\documentclass[conference]{IEEEtran} 
\IEEEoverridecommandlockouts

\usepackage{cite}
\usepackage[english]{babel}
\usepackage{caption}
\usepackage{tabularx}
\usepackage{mwe}
\usepackage{subcaption}
\makeatother

\usepackage{eso-pic}
\def\BibTeX{{\rm B\kern-.05em{\sc i\kern-.025em b}\kern-.08em
    T\kern-.1667em\lower.7ex\hbox{E}\kern-.125emX}}

\usepackage[linesnumbered,ruled,lined]{algorithm2e}
\usepackage{amsmath,amssymb,amsfonts}
\usepackage{algorithmic}
\usepackage{graphicx}
\usepackage{textcomp}   
\usepackage{xcolor}
\usepackage{nccmath}
\IEEEoverridecommandlockouts

\begin{document}
	
	\title{Indoor Positioning via Gradient Boosting Enhanced with Feature Augmentation using Deep Learning  \\
		\thanks{This paper is partly supported by Academy of Finland via (a) FIREMAN consortium n.326270 as part of CHIST-ERA grant CHIST-ERA-17-BDSI-003, and (b) EnergyNet Research Fellowship n.321265/n.328869 and (c) n.339541, and by Jane and Aatos Erkko Foundation via STREAM project.}
	}

\author{Ashkan Goharfar$ ^* $, Jaber Babaki$ ^{*} $, Mehdi Rasti$ ^{*} $, Pedro H. J. Nardelli$ ^\dagger $ \\
	$ ^* $Department of Computer Engineering, Amirkabir University of Technology, 
	Tehran, Iran\\
	$ ^\dagger $School of Energy Systems, {Lappeenranta-Lahti University of Technology}, Lappeenranta, Finland \\
	Email: {\{ashkan\_goharfar,babaki,rasti\}@aut.ac.ir}, pedro.nardelli@lut.fi}

	\maketitle

\IEEEpubidadjcol
\begin{abstract}
With the emerge of the Internet of Things (IoT), localization within indoor environments has become inevitable and has attracted a great deal of attention in recent years. Several efforts have been made to cope with the challenges of accurate positioning systems in the presence of signal interference. 
In this paper, we propose a novel deep learning approach through Gradient Boosting Enhanced with Step-Wise Feature Augmentation using Artificial Neural Network (AugBoost-ANN) for indoor localization applications as it trains over labeled data. For this purpose, we propose an IoT architecture using a star network topology to collect the Received Signal Strength Indicator (RSSI) of Bluetooth Low Energy (BLE) modules by means of a Raspberry Pi as an Access Point (AP) in an indoor environment. The dataset for the experiments is gathered in the real world in different periods to match the real environments. Next, we address the challenges of the AugBoost-ANN training which augments features in each iteration of making a decision tree using a deep neural network and the transfer learning technique. Experimental results show more than 8\% improvement in terms of accuracy in comparison with the existing gradient boosting and deep learning methods recently proposed in the literature, and our proposed model acquires a mean location accuracy of 0.77 m.
\end{abstract}


\section{Introduction}
Indoor localization is one of the most demanding applications in both business and public safety. Commercially, it can be used to track children, people with special needs, assist blind people in navigation, identify equipment, and mobile robots, among other things \cite{ref1}. Aside from making navigation easier for users, an indoor positioning system ensures a pleasant user experience and offers the option of heat mapping, which allows us to see how people move within a space.


Some of the indoor positioning systems are based on time-related data/estimation, such as Time of Arrival (ToA), Time of Flight (ToF), and Time Difference of Arrival (TDoA) \cite{ref2}, \cite{ref3}. Either appropriate time synchronization or an antenna array is required for the ToA, ToF, and TDoA positioning systems, which may raise the system cost. In contrast, a RSSI-based positioning system is based on the characteristics of wireless signal strength over time and does not require time synchronization or angle measurement \cite{ref4}. In RSSI-based positioning system, the AP collects the RSSI values from Reference Points (RPs) to make the fingerprint feature vectors of the location grids in indoor localization systems, known as the RSSI-based or fingerprinting datasets.

One of the main challenges of implementing the indoor positioning systems that use fingerprinting datasets is to find an appropriate supervised learning algorithm to classify locations based on their labels. In recent years, various kinds of shallow learning algorithms, for instance, k-Nearest Neighbors (k-NN) \cite{ref5}, Support Vector Machine (SVM) \cite{ref6}, Logistic Regression, Gradient Boosted Decision Tree (GBDT) \cite{ref16}, and Extreme Gradient Boosting (XGBoost) \cite{ref9} have been applied to the RSSI fingerprinting data. Despite the fact that, GBDT and XGBoost have achieved acceptable results, they could not reduce the significant effect of signal interference in the data which is one of the most challenging problems in indoor localization systems. Hence, some deep neural networks (DNNs) have been recently introduced to deal with the noise issue \cite{ref11, ref12, ref13}.

Deep learning is a convenient machine learning technique for feature augmentation algorithms which increase the statistical dependencies between the predictions of the individual base models \cite{ref10}. To elucidate higher level models of noisy inputs due to the signal fluctuations, the feature augmentation method extracts features from fingerprinting data. Transfer learning is a suitable machine learning approach for improving the feature augmentation algorithm's outcome. The idea behind transfer learning is to freeze the first few layers of a pre-trained Artificial Neural Network (ANN), then retrain the remainder of the layers on new data \cite{ref20}. Because the new task is comparable to the previous task, we presume that the embedding will be beneficial for the new task.

In this paper, we first gather the fingerprinting dataset in a BLE network. Then, we use a state-of-the-art algorithm named AugBoost-ANN which was introduced in \cite{ref10}. To the best of our knowledge, this study is the first to use the AugBoost-ANN for indoor positioning systems in support of IoT services.
Our contributions in this paper are summarized as follows:

\begin{itemize}
    \item We first prepare fingerprinting dataset by collecting the RSSI values from a few BLE nodes as RPs using a Raspberry Pi as an AP. Then, we propose an indoor positioning algorithm so called AugBoost-ANN, which 
    implements GBDT's stage-wise additive expansions with a neural-network-based feature augmentation method. Therefore, in each iteration of making a decision tree (DT) the algorithm combines the DT with an ANN.
    
    \item We compare our proposed technique with existing deep learning and gradient boosting algorithms recently proposed in literature, in terms of mean and standard deviation of location accuracy. 
    The mean accuracy of our proposed technique is 27\% , 70\%, and 19\% more accurate than \cite{ref11}, \cite{ref12}, and \cite{ref9}, respectively. Moreover, in terms of standard deviation,  the localization error of our proposed technique is 11\% better than that of \cite{ref9}.
\end{itemize}

The rest of this paper is organized as follows. In Section II, literature and background review are investigated. Our proposed indoor localization technique is discussed in detail in Section III. Section IV organizes experimental results for the indoor positioning system. Finally, Section V presents the conclusion and future works.

\section{BACKGROUND and LITERATURE REVIEW}
In this section, an overview of indoor localization is first provided and then the application of GBDT algorithm for indoor localization is discussed, and the related works are reviewed.

\subsection{An overview on Indoor Localization}
The process of monitoring an indoor place using data obtained from various sources such as wired or wireless networks is known as a indoor localization system \cite{ref7}. The majority of indoor localization systems have been suggested using Zigbee, Bluetooth, Wi-Fi, and cellular, with various degrees of implementation complexity and accuracy. BLE modules are more suitable for indoor localization, since 
it is a low-cost technology and  we need only to install  the battery-operated cheap BLE devices in the monitoring area.

\begin{figure}
    \centering
    \includegraphics[width=0.9\linewidth, height=4.0cm]{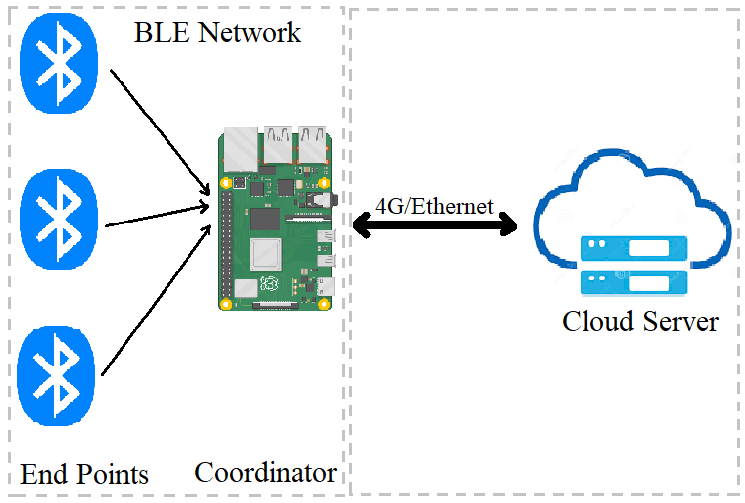}
    \caption{The general network architecture for the indoor positioning system.}
    \label{fig:BLE network artichectrue}
\end{figure}

A general architecture of an indoor positioning system consisting of BLE modules and a Raspberry Pi is presented in Fig. \ref{fig:BLE network artichectrue}. The presented network architecture consists of three main subsystems, including endpoints, coordinator, and cloud server. Endpoints consist of BLE modules located in different coordinates of the map provide RSSI data. The coordinator's main component is a Raspberry Pie that collects RSSI of endpoints, and it is in charge of communication between the cloud server and the endpoints. The coordinator is the center node of the star network topology, and it can gather and handle data without the need for contact with the cloud server. In the event that the server's communication link is broken (e.g. internet connection is down), the Raspberry Pi can realize the circumstance and will execute and train the learning model by itself \cite{ref15}. Also, the cloud server receives data from the coordinator and performs the training and stores the data through a database.

\subsection{Gradient Boosted Decision Tree}
 The concept of boosting arises from combining weak learners to get a model with significantly improved performance. Gradient Boosting is a machine learning approach to tackle regression and classification problems. It creates a prediction model using an ensemble of weak prediction models, such as decision trees \cite{ref16}. Bias Error and Variance Error are the two types of errors that can occur in machine learning systems. Gradient boosting is one of the boosting methods used to reduce the model's bias error. 
Gradient boosting includes three major components: a loss function, a weak learner, and an additive model. The loss function is responsible for optimization, and the weak learner's task is to make predictions. Moreover, the additive model is utilized for appending weak learners to reduce the loss value. 

The Gradient Boosted Decision Tree (GBDT) is a gradient boosting method based on decision trees that are comprised several decision trees in practical applications \cite{ref17}. The GBDT is based on a regression decision tree and is capable of adapting to non-linear features. It can be used to process various data like fingerprinting datasets. It has been successfully utilized in a variety of contexts because of its solid theoretical foundation, precise prediction, and simplicity. It is especially well suited to huge data applications, such as matrix and vector computations \cite{ref16}.

\subsection{Related Work}
Many of the indoor localization problems have either been solved or alleviated by the deployment of machine learning algorithms, however, each has its own strengths, deficiencies, and weaknesses. For example, various kinds of Recurrent Neural Networks (RNNs), including Long Short-Term Memory (LSTM), Gated Recurrent Unit (GRU), and bidirectional LSTM (BiLSTM), have been implemented and evaluated in \cite{ref13}. Despite that, they achieved a model with valuable mean location accuracy in terms of meter is presented in \cite{ref13}, the location accuracy standard deviation of the values obtained by their model is 0.64 m, which makes their indoor positioning system unreliable in some real indoor environments.

Some GBDT methods have been recently utilized to classify the locations based on the voting between decision trees models. For instance, in \cite{ref16}, the authors present a new multiple fingerprints method utilizing the GBDT and characteristics of RSSI. In \cite{ref18}, a Wi-Fi based indoor positioning system is proposed to obtain RSSI-based dataset, and an auto-encoder neural network is implemented as a feature extraction algorithm to deal with the noise in the fingerprinting dataset. Then, a novel GBDT algorithm so called LightGBM is used in \cite{ref18} to classify the locations.
Furthermore, a method for indoor positioning based on a semi-supervised deep reinforcement learning model is proposed in \cite{ref5}, which makes use of the enormous amount of unlabeled collected data. In 
\cite{ref14}, a deep learning technique is utilized to build a database by employing a channel measurement spatial beam signal-to-noise ratios (SNRs) as described in IEEE 802.11ad/ay standard.

\section{The Proposed Indoor Localization Technique}
\label{sec:proposed technique}
\begin{figure}
    \centering
    \includegraphics[width=1.0\linewidth, height=10.5cm]{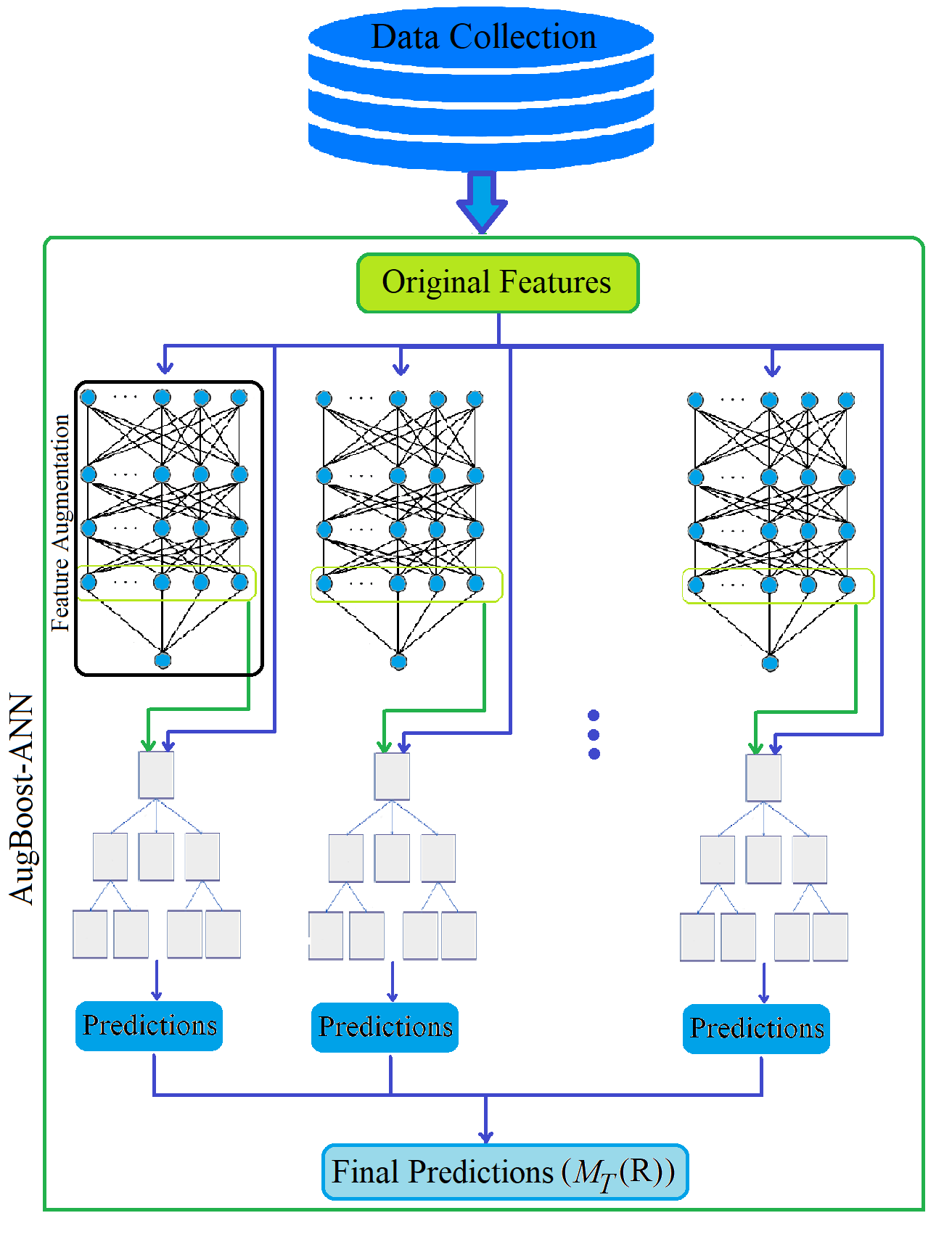}
    \caption{The general architecture of the proposed indoor localization technique.}
    \label{fig:augboost}
\end{figure}

In this section, we present our proposed indoor localization technique. We first describe the indoor environment for the data collection and the collected dataset. Then, the model's feature augmentation algorithm is presented which implements a deep neural network, followed by an introduction of AugBoost-ANN method for the indoor positioning system. The general architecture of the indoor localization technique is demonstrated in Fig. \ref{fig:augboost}.

\subsection{Data collection}
\begin{figure}
    \centering
    \includegraphics[width=1.0\linewidth, height=3.5cm]{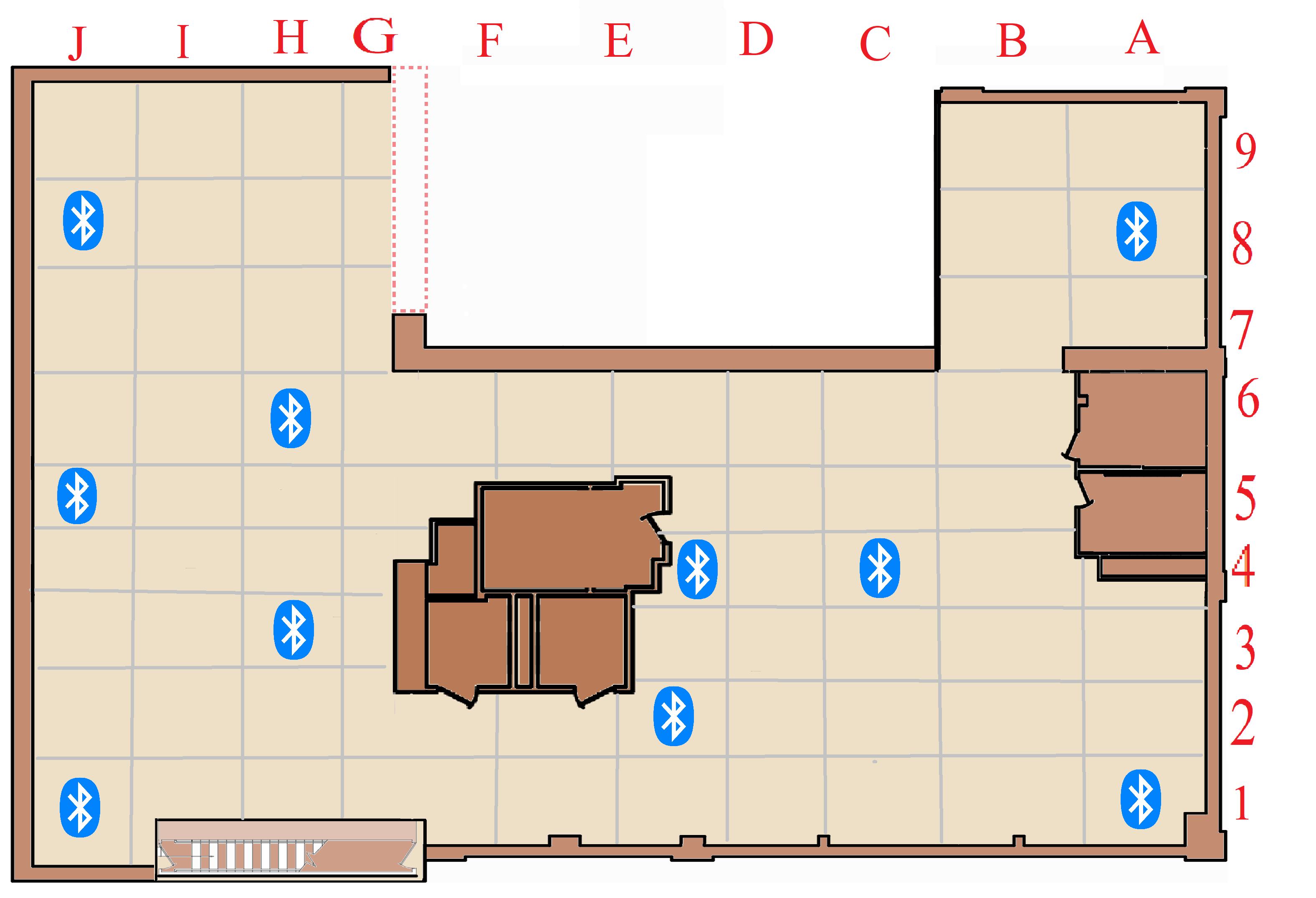}
    \caption{The illustration of a parking's map and 10 BLE nodes.}
    \label{fig:map}
\end{figure}

The dataset is collected in an indoor environment. At first, we partition the map into grids and then locate and fix a few BLE modules with a predetermined distance from each other. Then, we implement a scenario in which a Raspberry Pi is moving on different locations in different periods in the map, and collects data of BLE nodes every second, which are represented as iBeacon or Eddystone BLE profiles, and store RSSI parameters of transmitted packets in a dataset. The dataset is in CSV format and is illustrated with $m+1$ columns and $N$ rows, where $m$ in number of features (number of BLE nodes) and $N$ is the number of samples (seconds). Therefore, in the collected dataset, there are $m$ columns that contain RSSI of $m$ BLE modules and a column that shows the label of the location of Raspberry Pi for every second. Let denote the collected dataset by $S={(R_1,y_1),(R_2,y_2),(R_3,y_3),...,(R_N,y_N)}$, where $R_i=[{r_i^1},{r_i^2},{r_i^3},...,{r_i^m}]$  is the vector of collected m-dimensional RSSI, and $y$ is the vector of labels of the AP's locations. So, we have $N$ samples with $m$ features. Furthermore, we presume that $\left\{R_i\right\}_{i=0}^N$ is original features and $y$ is a target in our proposed technique.
An example of an indoor environment for a parking's map with 10 BLE modules is illustrated in Fig. \ref{fig:map}.

\subsection{Feature Augmentation with ANN}
\label{sec:feature augmentation}

The original features do not change before each iteration of making a Decision Tree (DT) during the training phase of the GBDT. Therefore, for indoor positioning tasks, the GBDT will not succeed in achieving an accurate model to classify the locations in the presence of the signal interference. Feature augmentation as a common technique for tackling Multi-Dimensional Classification (MDC) issues, manipulates the feature space by incorporating the label information \cite{ref21}. According to \cite{ref10}, a state-of-the-art method so called feature augmentation with Artificial Neural Network (ANN) has been proposed, which aims to train an ANN until the loss stops to improve, with the original features and the updated target. The ANNs architecture includes three fully connected hidden layers with Rectified Linear Unit ($ReLU$) activation function for each hidden layer which is defined as:

\begin{equation}
\label{eq1}
    Relu(x) = {max\ (0,x)}.
\end{equation}

\begin{figure}
    \centering
    \includegraphics[width=0.9\linewidth, height=4.9cm]{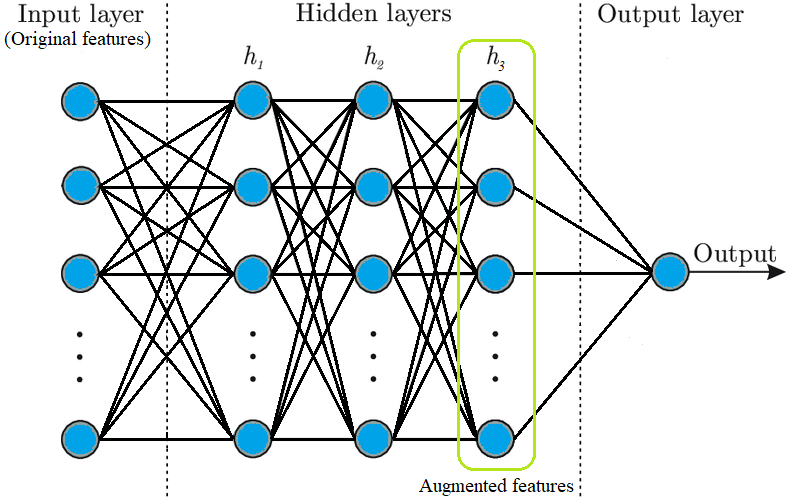}
    \caption{Representation of the proposed ANN structure.}
    \label{fig:ANN}
\end{figure}

The number of neurons in each hidden layer is equal to the number of input neurons. Furthermore, the batch size value is between 300 and $\frac{1}{15}$ of the samples in the dataset. In order to extract features from the ANN, neurons of the 3rd hidden layer are considered the augmented features using the transfer learning technique. Thus, we freeze the first few layers of a trained ANN and retrain the remainder of the layers on new data \cite{ref20}. Since the new task is comparable to the previous task, we presume that the embedding will be beneficial for the new task. This is the case when both tasks are similar, and it is thus prudent to retain as many layers as feasible from the pretrained ANN. This implies that we just dump the ANN's final layer \cite{ref10}. The structure of the proposed feature augmentation with ANN is presented in Fig. \ref{fig:ANN}.

\subsection{Indoor localization via AugBoost-ANN}
\label{sec:augboost}

As mentioned in the previous subsection, one of the main weaknesses of GBDT is that the original features do not change before each iteration. Therefore, feature augmentation algorithms have been proposed to increase the accuracy of the model in case of having signal fluctuation in the data. As Fig. \ref{fig:augboost} has shown, the gradient boosting enhanced with step-wise feature augmentation using artificial neural network (AugBoost-ANN) uses the proposed ANN as feature augmentation algorithm for the fingerprinting data before each iteration of creating a DT. At the end, AugBoost-ANN votes between predictions of all DTs with their predetermined weights to find the best model. The main idea of AugBoost-ANN was first proposed in \cite{ref10}, and organized the training procedure in Algorithm \ref{Alg:augboost}. 

\newcommand\mycommfont[1]{\footnotesize\ttfamily\textcolor{blue}{#1}}
\SetCommentSty{mycommfont}

\SetKwInput{KwInput}{Input}                
\SetKwInput{KwOutput}{Output}              




\begin{algorithm}
\caption{AugBoost-ANN Training Procedure}
\label{Alg:augboost}
\SetAlgoLined
\textbf{Input}:
\begin{small}
$S=\{\left.\begin{aligned}(R_1,y_1),(R_2,y_2),(R_3,y_3),...,(R_N,y_N)\end{aligned}\right\}$ \\
 \end{small}
\textbf{Output}: $M_T(R)$ \\
$M_0(R)=\underset{\rho}{\arg\min} \sum_{i=1}^N (\mathcal{L}(y_i, \rho))$ \\
\For{$i=1:N$}{
    Initialize the target for each sample: ${y}_{i}={y}_{i}$}
\For{$t=1:T$}{
    \uIf{$t-1$ is divisible by $c_{BA}$}{
        Split features of $R_i$ for district $i \in \{1,\dotsc,N\}$ to J random subsets. \\
        \For{$j=1:J$}{
        Apply feature augmentation method to $j^{\text{\tiny th}}$ subset using the proposed ANN which retrains the model for the subset and is represented in Fig 
        \ref{fig:ANN}.
        }
        Finally, extract augmented features from $3$rd hidden layer of the proposed ANN. \\
      }
      \Else{
        Set the feature augmentation method and subsets similar to them in the previous iteration. \\
    }
\For{$i=1:N$}{
	obtain $LGL$ from (\ref{eq3}) and update the target ${y}_{i}: \; {y}_{i} =LGL$}
Train $D_t$ through the augmented features and targets. Set values of the weights ($\rho_t$) using (\ref{eq4})\\
$\mathit{M_t(R)}=M_{t-1}(R)+\rho_t.D_t(R')$
}
\end{algorithm}

The algorithm takes $S$ as an input and returns a model named $M_T(R)$, which is based on a vote between predictions of $T$ decision trees with their corresponding weights. The algorithm defines $\mathcal{L}$ as a loss function which is mean square error ($\mathit{MSE}$), and is presented as follows,

\begin{equation}
\label{eq2}
\mathit{MSE}=\frac{1}{n} \sum_{i=1}^n (z_{i}-\hat{z}_{i})^2, 
\end{equation}

\noindent where $n$ is the number of data points, $z_{i}$ is observed values, and $\hat{z}_{i}$ is predicted values. In the first stage of training, we initialize $M_{0}(R)$ with a value of $\rho$ for which loss functions for $\tilde{y}_{i}$ and $\rho$ attains its minimum. Moreover, $\tilde{y}_{i}$ for distinct $i \in \{1,\dotsc,N\}$ are the targets for the algorithm, and before starting iterations, we initialize each of the $\tilde{y}_{i}$ with $y_{i}$ as early targets. The main purpose of GBDT is to make a decision tree in each of $T$ iterations and vote between predictions of $T$ decision trees with their predetermined weights to find the best model. 

The entire algorithm of AugBoost-ANN is similar to the GBDT algorithm except for the feature augmentation part, which should be taken place in each iteration before creating a decision tree. Although, We do not train the model used for feature augmentation in all iterations of making decision trees. Instead, starting with the first iteration, we retrain the model for every $c_{\textrm {BA}}$ iterations (BA is an acronym for 'Between Augmentations'). The model from the previous iteration is replicated in the subsequent iterations. Because each decision tree may only be able to use a portion of the data in each set of new features, this is designed to allow the boosting process to exploit the information in each set of new features. Therefore, for $t^{\text{\tiny th}}$ iteration, if $t–1$ is divisible by $c_{\textrm {BA}}$ we split features of $R_{i}$ for distinct $i \in \{1,\dotsc,N\}$ to $J$ random subsets, and we apply the feature augmentation algorithm to each subset by means of the transfer learning technique. On the other hand, if $t–1$ is not divisible by $c_{BA}$ we set the feature augmentation method and subsets similar to them in the previous iteration. The reason behind using $\mathit{MSE}$ as a loss function is to measure the quality of each split.

Next, we update the targets to assume the ensemble's errors from the previous iteration using the last gradient of the loss function denoted by $LGL$ for each sample of the original features, which is given as:

\begin{equation}
\label{eq3}
 \mathit{LGL}=-\left[ \frac{\partial }{\partial M(R_i)}{\mathcal{L}(y_i, M(R_i))} \right]_{M(R)=M_{t-1}(R)}.
\end{equation}

Afterward, we train a decision tree using the output of the feature augmentation algorithm and the updated targets. 
We set the weight of the model which is generated from the decision tree in $t^{\text{\tiny th}}$ iteration and denoted by $\rho_t$  for optimizing loss function as:

\begin{equation}
\label{eq4}
 \mathit{\rho_t}=\underset{\rho}{\arg\min} \sum_{i=1}^N (\mathcal{L}(y_i, M_{t-1}(R_i)+\rho.D_t(R_i^\prime))),
\end{equation}
where $D_t$  is the decision tree in $t^{\text{\tiny th}}$ iteration, $M_{t-1}(R)$ denotes the ensemble of models generated in the previous iteration, and $R_i^\prime$ is the $i^{\text{\tiny th}}$ augmented feature that gives to $D_t$ as an input. Hence, we initialize $\rho_t$ with a value of $\rho$ for which loss functions attains its minimum. After all, $M_{t}(R)$ is obtained from ensembles of previous models.


\section{EXPERIMENTAL RESULTS}
\label{sec:result}
In this section, we evaluate feature augmentation and supervised learning methods in terms of location accuracy for the different iterations. At first, we collect the fingerprinting dataset in a parking's of a building with a size of $12.5\times18$ square meters and partition the map into grids of size $1.25\times1.5$ square meters. Next, we locate 10 BLE modules ($m=10$) with the distance of approximately 4 meters. Finally, we move a Raspberry Pi 4 B+ in the map during 1090 seconds ($N=1090$) in different periods to gather the data. The full document and dataset are accessible in \cite{ref19}.

For our experiments, all of the results are represented as outcomes for the calculation of mean accuracy and standard deviation in terms of meters using Euclidean distance for locations. The standard deviation formula is defined as:

\begin{equation}
\label{eq5}
 \sigma=\sqrt{\frac{(\sum(x_i-\mu))^2}{n}},
\end{equation}

\noindent where $\sigma$ is accuracy standard deviation, $n$ is the number of the training processes, $x_i$ is a value from the list of all accuracies in all training processes, and $\mu$ is the mean accuracy.

\subsection{Feature Augmentation}
A feature augmentation algorithm so called Random Projection (RP) is suggested in \cite{ref10}, where instead of projecting all of the features with the same random projection, the algorithm applies the projections independently to each of the randomly chosen feature subsets. Then, rather than simply employing the new features, it concatenates the original and new features. After all, per number of iterations, it just re-extracts the features once \cite{ref10}. We compare an appropriate use of ANN approach to enhance the outcomes from utilizing RP and to augment the most important features in the algorithm that played a significant role in an accurate location classifier as one of the primary contributions in the suggested indoor localization methodology. For this purpose, we apply AugBoost-ANN, AugBoost-RP, and XGBoost on one fold test with 70\% of the dataset random training trajectories samples. 

\begin{figure}
    \centering
    \includegraphics[width=0.8\linewidth, height=4.2cm]{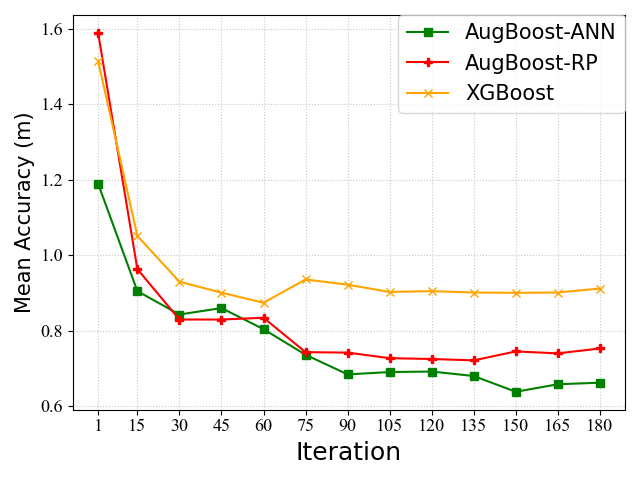}
    \caption{The comparison of mean accuracy (m) of AugBoost-ANN, AugBoost-RP, and XGBoost on one fold for different iterations.}
    \label{fig:ANN_vs_RP}
\end{figure}

Fig. \ref{fig:ANN_vs_RP} shows mean accuracy of AugBoost algorithm using ANN and RP for the different number of iterations. It is noticed that AugBoost-ANN is more accurate than AugBoost-RP in 85\% of the iterations. Especially, AugBoost-ANN achieves its best mean accuracy of 0.64 meters in $150^{\text{\tiny th}}$ iteration; however, AugBoost-RP obtains the mean accuracy of 0.72 meters in $135^{\text{\tiny th}}$ iteration.

XGBoost is one of the most common gradient boosting methods, and also belongs to GBDT family, which is implemented in \cite{ref9} for indoor positioning systems and outperforms some of shallow learning algorithms in terms of mean accuracy. As Fig. \ref{fig:ANN_vs_RP} has shown, AugBoost-ANN outperforms XGBoost with regard to mean accuracy in all iterations. Although, the mean accuracy of XGBoost is better than AugBoost-RP in the first few iterations.

\subsection{Localization Performance}
As the second experiment, we evaluate location accuracy in terms of mean accuracy and standard deviations for different supervised learning algorithms. For implementing AugBoost-ANN algorithm, we use 150 iterations and $c_{BA}$ equal to 15, hence, all over the procedure, we train 150 DTs and 15 times retrain the proposed ANN. Moreover, we choose Adam optimizer as an optimization function and set the number of epochs and learning rate to 30 and 0.1, respectively. All of the results are presented after 8-fold tests with 70\% of the dataset random training trajectories samples. In our research, the Keras package is utilized for implementing the deep neural network on TensorFlow \cite{ref22}. To evaluate our proposed deep learning approach, we compare some supervised learning algorithms which were presented in the recent publications in the field of indoor localization systems and are demonstrated in Table \ref{table:comparison}.

\begin{table}
\begin{center}
\caption{\label{table:comparison}The comparison of our proposed technique after 8-fold tests with the existing related works.}
\begin{tabular}{||c c c c||} 
\hline
 Algorithm & labels & Grid size & Location accuracy \\ [0.3ex] 
 \hline\hline
 MLP \cite{ref11} & 50 & 1.7 m & $2.8 \pm 0.1$ m \\ [0.3ex]
 \hline
 MLNN \cite{ref12} & 20 & 1.5 m & $1.1 \pm 1.2$ m \\ [0.3ex]
 \hline
 XGBoost \cite{ref9} & 1401 & 1.5 m & $3.99 \pm 2.81$ m \\ [0.3ex]
 \hline
 Proposed AugBoost-ANN & 54 & 1.25-1.5 m & $0.77 \pm 0.3$ m \\ [0.3ex]
 \hline
\end{tabular}
\end{center}
\end{table}

According to Table \ref{table:comparison}, we compare our proposed method with deep learning algorithms named Multi-Layer Perceptron (MLP) \cite{ref11}, Multi-Layer Neural Network (MLNN) \cite{ref12}, and Extreme Gradient Boosting (XGBoost) \cite{ref9} that each of which has their strength and weaknesses. For example, MLP obtains a standard deviation of 0.1 m which is 0.2 m better that the standard deviation of AugBoost-ANN. However, its mean accuracy is 2.8 m which means that our proposed method is more reliable in this matter. Also, despite that, XGboost and AugBoost-ANN are based on GBDT, XGBoost achieved a mean accuracy and standard deviation of 3.99 m and 2.81 m, which indicates that applying the feature augmentation method improves accuracy significantly. Furthermore, compared with MLNN, AugBoost-ANN has performed better in all respects.

\begin{figure}
    \centering
    \includegraphics[width=0.8\linewidth, height=4.2cm]{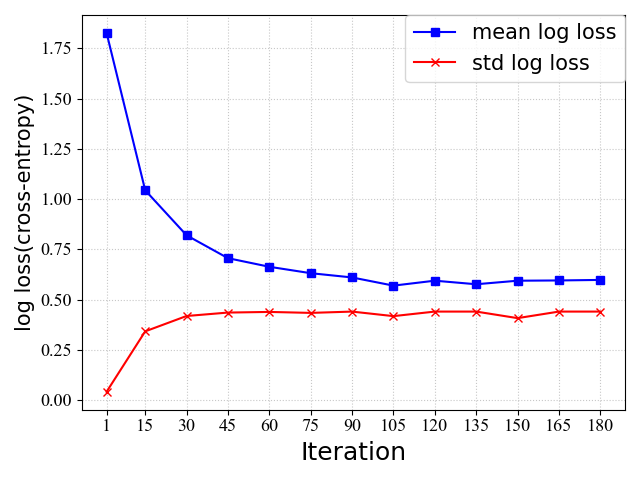}
    \caption{The learning curve of AugBoost-ANN on one fold for different iterations with respect to the log loss.}
    \label{fig:mea_log_std}
\end{figure}

Finally, we investigate the performance of the deep neural network to evaluate the mean and standard deviation (std) of logarithm with base 10 of the loss function (log loss) for the different number of iterations. The loss function is cross-entropy since our neural network augments features. Also, from Fig. \ref{fig:mea_log_std}, we observe that from the first iteration to $45^{\text{\tiny th}}$ iteration mean log loss decreased rapidly from 1.8 to 0.7, and std log loss raise significantly by 0.5. From $45^{\text{\tiny th}}$ iteration to $180^{\text{\tiny th}}$ iteration, these values change smoothly and being stabled in $150^{\text{\tiny th}}$ iteration. The code repository is accessible in \cite{ref23}.

\section{conclusion}
\label{sec:conclusion}
In this paper, we proposed a novel deep learning solution for classifying locations in an indoor environment using BLE networks. The current study is the first to use AugBoost-ANN for indoor positioning systems. For this purpose, we utilized an IoT architecture with star network topology with 10 BLE modules and a Raspberry Pi to gather fingerprinting data. Then, we used a novel deep learning approach called AugBoost-ANN, which augments features in each iteration of making a decision tree using a deep neural network and transfer learning technique. Our proposed algorithm outperforms the gradient boosting and deep learning algorithms for indoor localization, recently proposed in related works. As our future work, we will investigate new deep learning approaches such as Auto Encoder networks for feature extraction. 


\bibliographystyle{IEEEtran}
\bibliography{Mybib}

\end{document}